# ACTIVE ROAD NOISE CONTROL USING NETWORKED MEAN-SQUARE LEARNING AND LIGHTWEIGHT CONDITIONAL COMPENSATION EXPERTS

*Xingzhao Huang[1], Hengzhi Zhou[1], Yu Shi[1], Ming Wu[2], Bin Liao[3], Yijing Chu[1]*

[1] School of Architecture, South China University of Technology, Guangzhou 510641, China
[2] Institute of Acoustics, Chinese Academy of Sciences, Beijing 100190, China
[3] College of Electronics and Information Engineering, Shenzhen University, Shenzhen 518060, China

## ABSTRACT

Multichannel active road noise control (ARNC) requires effective noise reduction (NR) under causality and computational constraints. A computationally efficient method is proposed that separates linear control from condition-dependent nonlinear residual compensation. A finite impulse response (FIR) baseline filter is obtained through networked mean-square learning that can be made adaptive, while a multi-resolution FIR mixture-of-experts (MRF-MoE) module provides possibly nonlinear corrections. Training comprises baseline initialization, sample-graph-regularized expert and router optimization, and joint refinement. Target channels are treated as nodes in a connected network for mean-square error minimization, whereas training windows define a graph for regularizing routing probabilities. Each expert combines short-time dense and long-tail low-rank FIR filtering, and causal single-expert routing reduces inference computation. Compared with WaveNet-VNN under the same configuration, the proposed method reduces estimated inference cost by 96.3% while achieving higher NR on the test set.

## 1. INTRODUCTION

Active road noise control (ARNC) suppresses low-frequency cabin noise through the anti-noise generated by secondary sources [1–3]. Most systems adopt the multichannel adaptive filtering framework of active noise control (ANC), using filtered-x least mean square (FxLMS) algorithms [2] and their variants [4–8]. These methods achieve a trade-off between computational cost, convergence, and noise attenuation [9].

Nonlinear vibroacoustic transmission and changing operating conditions further limit linear adaptive control. Suspension nonlinearities can weaken reference–noise coherence [10], while acoustic changes can cause mismatch in secondary-path models and remote microphone technique (RMT) observation filters [11,12]. Control performance is thus constrained by the linear controller's representational capacity and the accuracy of the acoustic models.

Recent neural approaches learn nonlinear control mappings offline and retain fixed parameters during inference. Deep MCANC provides end-to-end multichannel control [13], whereas WaveNet–Volterra Neural Network (WaveNet-VNN) and MVNet combine Volterra filtering with causal temporal modeling [14,15]. Selective fixed-filter ANC (SFANC) enables learned filter selection [16,17]. Feedback-guided deep neural network (DNN)-based controller fusion [18], referred to here as FG-DNN, combines FIR experts with a neural controller.

However, deep controllers require substantial inference computation, and offline-trained controllers may lose performance when road conditions or vehicle speed change [19]. Adaptive WaveNet-VNN updates lightweight adapters online [19]. Secondary-path learning, joint controller–path modeling, and meta-learned initialization support adaptation to acoustic changes [11,20,21]. Hybrid SFANC with filtered-x normalized least mean square (FxNLMS) adaptation selects pretrained filters to initialize subsequent iterations [22]. These approaches motivate a computationally efficient FIR-based structure that combines multichannel linear filtering with condition-dependent nonlinear compensation.

We propose a multichannel ARNC controller that combines a baseline FIR, compensation experts, and a causal router. Baseline-filter initialization is formulated as a networked mean-square error (MSE) minimization problem over a connected network of target-channel nodes [23,24]. A sample graph is constructed from training windows to regularize routing probabilities for similar signal histories during expert and router training [25]. Each expert combines short-time dense and long-tail low-rank FIR filtering, while single-expert sparse execution reduces inference computation. The FIR-based structure also provides a basis for future integration with adaptive filtering.

## 2. PROPOSED METHOD

### 2.1. Problem formulation and optimization strategy

ARNC seeks to minimize road noise at target positions under causality constraints. The controllers in Fig. 1 generate $\mathbf{y}(n)$ from acceleration references $\mathbf{x}(n)$ and physical microphone observations. Propagation through the secondary paths $\mathbf{S}_v$ gives the virtual-target error $\mathbf{e}_v(n)$. At the physical microphones, the measured error satisfies $\mathbf{e}_p = \mathbf{d}_p + \mathbf{S}_p * \mathbf{y}$ ; subtracting the estimated secondary-path response gives the primary-noise estimate $\hat{\mathbf{d}}_p = \mathbf{e}_p - \hat{\mathbf{S}}_p * \mathbf{y}$ . Acceleration references and estimated primary noise are concatenated as $\tilde{\mathbf{x}} = \left[\mathbf{x}, \hat{\mathbf{d}}_p\right]$. Primary paths to the physical and virtual target positions are denoted by $\mathbf{P}_p$ and $\mathbf{P}_v$, respectively.

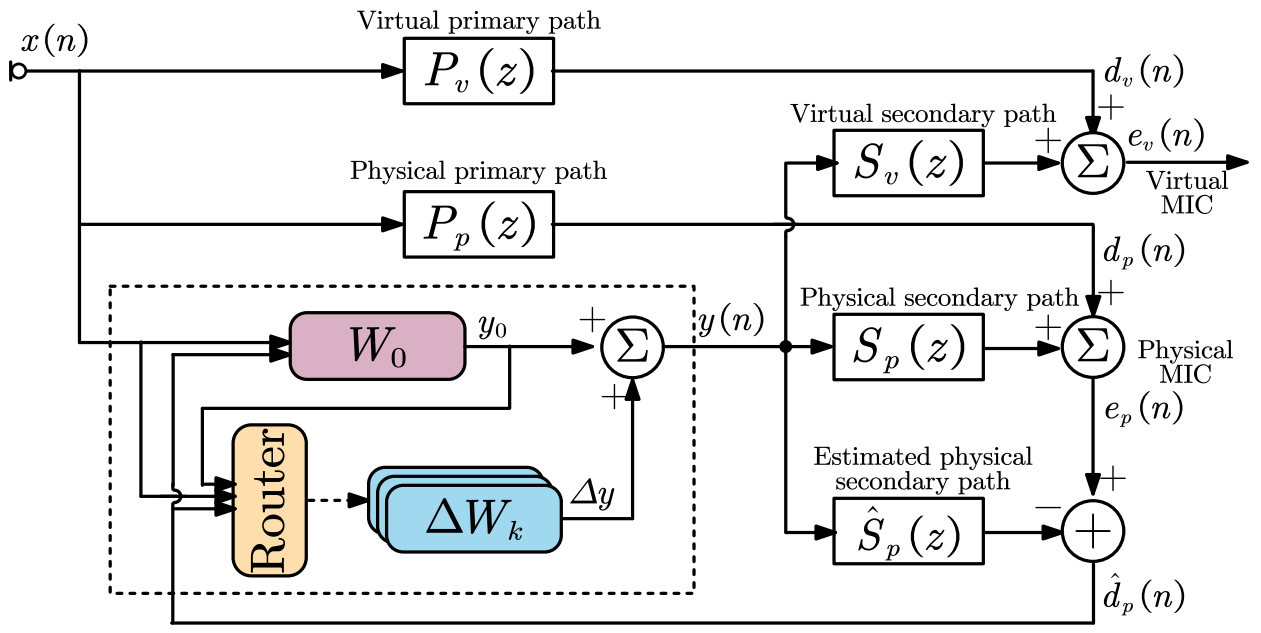

**Fig. 1.** Block diagram of the proposed method.

The proposed method consists of a baseline FIR filter $\mathbf{W}_0$ and an MRF-MoE module comprising compensation experts $\{\Delta\mathbf{W}_k\}$ with a causal router (Fig. 1). The baseline filter generates anti-sound for linear control in the mean-square sense, while the selected expert produces a condition-dependent signal for local nonlinear compensation. Both branches produce control signals directly through FIR filtering.

Controller learning is formulated over a graph of condition-specific datasets and a sample graph. For baseline learning, the connected graph $\mathcal{G} = (\mathcal{V}, \mathcal{C})$ contains one logical node per operating condition [23,24]. Each node contributes a channel-weighted MSE evaluated on its condition-specific multichannel data, and the equality constraints enforce a common baseline across operating conditions. The sample graph regularizes routing probabilities for similar signal histories [25]. Training comprises baseline initialization, expert and router optimization, and joint refinement.

The control output is $\mathbf{y} = \mathbf{y}_0 + \Delta\mathbf{y}_{k_b}$ , combining the baseline multiple-input multiple-output (MIMO) FIR response with the selected expert's compensation signal.

The error after baseline filtering is $\mathbf{e}_{v,0} = \mathbf{d}_v + \mathbf{S}_v * (\mathbf{W}_0 * \tilde{\mathbf{x}})$. Each expert provides a compensation signal $\Delta\mathbf{y}_k$. The routed compensation yields the target error $\Delta\mathbf{e}_v = \mathbf{e}_{v,0} + \mathbf{S}_v * \Delta\mathbf{y}_{k_b}$ . Baseline and compensation learning are formulated within the following unified optimization problem:

$$\min_{\{\mathbf{w}_{0,i}\},\vartheta,\psi} \sum_i \gamma_i \mathbb{E}_i \left[ \left\| \mathbf{e}_{v,0,i}\left(\mathbf{w}_{0,i}\right) \right\|_{\mathbf{P}_i}^2 \right] + \left[ \mathcal{L}_c\left(\Delta\mathbf{e}_v(\vartheta)\right) + \lambda_G \mathcal{R}_G(\psi) \right] \quad (1)$$
$$\text{s.t.} \quad \mathbf{w}_{0,i} = \mathbf{w}_{0,j}, (i,j) \in \mathcal{C}.$$

Here, $i$ indexes operating conditions, $\gamma_i$ weights the corresponding condition-specific objective, and $\mathbb{E}_i$ denotes averaging over observations under that condition. The vector $\mathbf{e}_{v,0,i}$ collects the baseline errors at all target channels, and $\mathbf{P}_i$ specifies their relative weights. Each $\mathbf{w}_{0,i}$ is a local copy of the baseline FIR coefficient vector, and the equality constraints enforce a common baseline across operating conditions. The expert and router parameters are denoted by $\vartheta$ and $\psi$, respectively. The second term combines the control loss $\mathcal{L}_c$ with the sample-graph regularizer $\mathcal{R}_G$.

Baseline-filter initialization and compensation learning are organized into a three-stage optimization procedure, combining mean-square learning, sample-graph regularization, and joint refinement:

1) Networked baseline initialization: The baseline FIR is initialized by minimizing a regularized version of the first term in Eq. (1), combining condition-weighted objectives and channel-weighted errors under the equality constraints.

2) Graph-regularized compensation learning: With the baseline FIR fixed, the MRF-MoE module is optimized using the second term in Eq. (1), which combines the control loss with sample-graph regularization of routing probabilities [25].

3) Joint refinement: The baseline FIR is unfrozen, and all controller components are jointly optimized to reduce in-band noise and suppress out-of-band (OOB) amplification while penalizing excessive output. The baseline-only and sample-graph terms in (1) are omitted.

Steps 2 and 3 can be made adaptive in nonstationary environments and the MRF-MoE has nonlinear modeling capability.

## 2.2. MRF compensation experts and causal routing

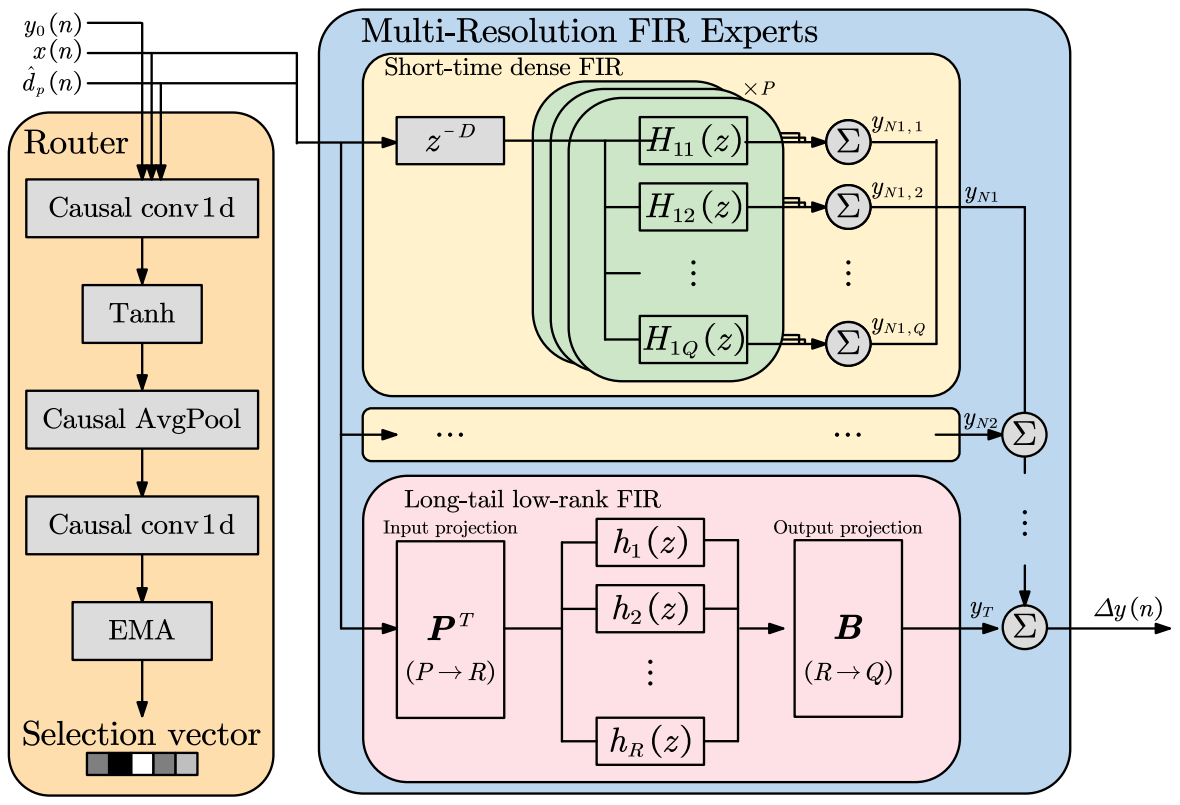

**Fig. 2.** Architecture of the causal router and MRF experts.

The MRF-MoE module in Fig. 2 contains $K$ compensation experts with identical architectures and independent parameters. Each expert combines one or more delayed short-time dense FIR branches with a long-tail low-rank FIR branch. All branches are causal, and the expert output is

$$\Delta \mathbf{y}_k = \sum_l \mathbf{A}_{k,l} * \left(z^{-D_l} * \tilde{\mathbf{x}}\right) + \mathbf{B}_k\left[\mathbf{h}_k *_{\mathrm{dw}} \left(\mathbf{P}_k^{\mathrm{T}}\tilde{\mathbf{x}}\right)\right]. \quad (2)$$

Here, $*$ denotes causal MIMO convolution, and $*_{\mathrm{dw}}$ denotes depthwise causal convolution. The dense filter $\mathbf{A}_{k,l}$ models the short-time response of the $l$-th dense branch in expert $k$, with a nonnegative input delay $D_l$. Multiple dense branches can use different delays to capture local responses at selected lags. The evaluated configuration uses a single dense branch with $D_1 = 0$. The long-tail low-rank branch comprises input projection $\mathbf{P}_k^{\mathrm{T}}$, latent-channel FIR $\mathbf{h}_k$, and output mixing $\mathbf{B}_k$. The same projection and mixing matrices are used across filter taps, extending the modeled history with fewer parameters.

The router receives the controller input and baseline-filter output and selects a compensation expert using features associated with the current operating condition. Channelwise normalization is followed by a causal convolution, Tanh activation, causal average pooling, and a dilated causal convolution. The network generates routing scores $\mathbf{s}_b$ at the start of each control block.

An exponential moving average (EMA) smooths the scores. Scores from the preceding block determine the current expert:

$$\begin{aligned} \bar{\mathbf{s}}_b &= \rho\bar{\mathbf{s}}_{b-1} + (1-\rho)\mathbf{s}_b, \\ \pi_b &= \mathrm{softmax}\left(\bar{\mathbf{s}}_{b-1}/\tau\right), \\ k_b &= \arg\max_k \pi_{b,k}. \end{aligned} \quad (3)$$

Here, $\rho$ is the smoothing coefficient and $\tau$ is the softmax temperature. The forward pass selects the expert with the largest probability, whereas the backward pass uses the soft probabilities for straight-through gradient estimation. The selected expert remains fixed within each control block.

### 2.3. Staged optimization strategy

In the first stage, the baseline FIR is initialized using the causal Wiener–Hopf solution of a regularized version of the first term in (1). Here, $\mathbf{\Phi}_i$ stacks the secondary-path-filtered regressors for all target channels, and $\mathbf{d}_{v,i}$ contains the corresponding primary disturbances. With $\mathbf{H}_i = \mathbb{E}_i\left[\mathbf{\Phi}_i^{\mathrm{T}}\mathbf{P}_i\mathbf{\Phi}_i\right]$ and $\mathbf{p}_i = \mathbb{E}_i\left[\mathbf{\Phi}_i^{\mathrm{T}}\mathbf{P}_i\mathbf{d}_{v,i}\right]$, the initialization satisfies

$$\left(\sum_i \gamma_i\left(\mathbf{H}_i + \mathbf{R}_i\right)\right)\mathbf{w}_0^{\mathrm{init}} = -\sum_i \gamma_i \mathbf{p}_i. \quad (4)$$

Here, $\mathbf{R}_i$ is a positive semidefinite matrix specifying control-energy regularization and numerical ridge stabilization. The condition-weighted statistics in (4) yield a common multichannel Wiener baseline. The evaluated model starts from this baseline and a pretrained MRF-MoE module.

In the second stage, the MRF-MoE module is optimized with the baseline FIR fixed, making the first term in (1) constant. Training windows are treated as nodes of a fixed sample graph, with edges defined by similarities between the observable histories preceding control decisions. The graph regularizer penalizes differences in routing probabilities between connected nodes [25]:

$$\min_{\vartheta,\psi} \mathcal{L}_{\mathrm{c}} + \frac{\lambda_G}{Z_d}\mathrm{tr}\left(\mathbf{\Gamma}_\psi{}^{\mathrm{T}}\mathbf{L}_d\mathbf{\Gamma}_\psi\right). \quad (5)$$

Here, the rows of $\mathbf{\Gamma}_\psi$ are the router probabilities at the sampled decision points. $\mathbf{L}_d$ is the unnormalized Laplacian of the undirected sample graph, $Z_d$ sums its edge weights once per edge, and $\lambda_G$ controls graph regularization. The normalized trace defines $\mathcal{R}_G$ in (1). The experts and router are optimized by backpropagation, with the graph term acting directly on the routing probabilities.

In the third stage, all controller parameters are updated using only the control loss, omitting the baseline-only and sample-graph terms of (1). The control loss used in (1) and (5) is

$$\mathcal{L}_{\mathrm{c}} = 10\log_{10}\frac{P_A\left(\Delta e_{v,j,m};\mathcal{F}\right)}{P_A\left(d_{v,j,m};\mathcal{F}\right)} + \lambda\mathcal{L}_{\mathrm{OOB}} + \lambda_y\mathcal{L}_y. \quad (6)$$

Here, $P_A(q;\mathcal{F})$ denotes full-recording A-weighted power in the target band; $j$ and $m$ index recordings and target channels, respectively. The first term is the negative A-weighted NR for each recording and target channel. Recording and target-channel averaging is implicit in $\mathcal{L}_{\mathrm{c}}$. The spectral power change $r_{j,m}(f)$ is the ratio of error power to primary-noise power, expressed in decibels, calculated from unweighted Welch spectra; positive values indicate amplification. The OOB penalty $\mathcal{L}_{\mathrm{OOB}}$ averages the largest squared positive exceedances above $\delta$, retaining at most $J$ entries across target channels and OOB frequency bins per recording. It is zero when there are no exceedances. $\lambda$ controls OOB penalty strength. The output penalty $\mathcal{L}_y$ averages squared relative exceedances above the normalized amplitude limit of 0.95 over the largest 0.1% of samples per output channel. Its weight is $\lambda_y$ = 1. Secondary paths and normalization statistics remain fixed throughout training.

## 3. EXPERIMENTAL EVALUATION

### 3.1. Experimental setup

Experiments use the vehicle dataset for the automotive road-noise cancellation task of the 2026 Fourth Shenghua Cup Acoustics Technology Competition [26]. It contains 35 training and 17 public test recordings covering Mixed, Rough, and Smooth roads at 40 and 60 km/h, sampled at 2 kHz. The system uses 12 acceleration references, four physical microphones, six secondary sources, and two target positions at the driver's ears.

The proposed method uses a 64-tap baseline FIR and an MRF-MoE module with four experts, each combining an 80-tap dense branch with a 256-tap low-rank branch with eight latent channels. The router updates every 32 samples (16 ms), with an EMA coefficient of 0.995 and a softmax temperature of 0.2. Training follows Section 2.3, using an eight-nearest-neighbor sample graph constructed from log-band-power features and a graph weight of 0.3. For the OOB penalty, we set $\delta = 1\text{dB}$ , $\lambda = 0.2$ , and $J = 16$ .

Baselines include trained 256- and 512-tap FIR filters initialized from Wiener solutions, FxLMS [2] with a step size of 0.3, WaveNet-VNN [14] and FG-DNN [18]. All neural baselines use the same sensor configuration. Mean A-weighted NR over 30–350 Hz is averaged across test recordings and target channels. Maximum OOB amplification is evaluated across recordings and target channels over 20–1000 Hz, excluding the target band. Computational cost is estimated in thousands of multiply–accumulate operations per multichannel sample (kMAC/sample), including FxLMS updates and assuming single-expert sparse execution for the proposed method.

### 3.2. Results and discussion

The proposed method achieved a mean A-weighted NR of 7.53 dB(A) over the 30–350 Hz target band. The spectral curves in Fig. 3 show the spectral NR, with shaded envelopes indicating variability across the test recordings.

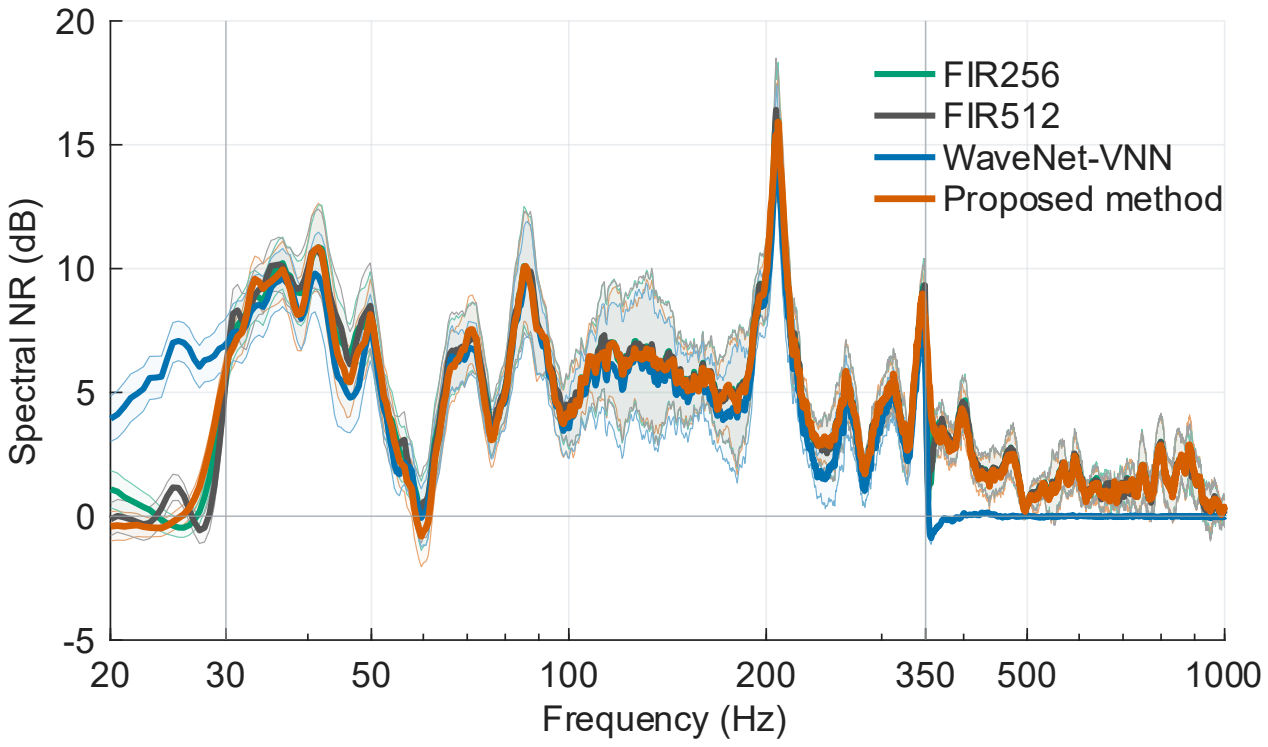

**Fig. 3.** Spectral NR on the test set. Thick curves show means; envelopes indicate ±1 sample standard deviation across recordings, smoothed over nine frequency bins.

Compared with WaveNet-VNN under the same channel configuration [14], the proposed method achieved slightly higher mean NR (Table 1). Its estimated inference cost decreased from 493.6 to 18.1 kMAC/sample, a reduction of 96.3%. Relative to the 256-tap FIR, the proposed method achieved comparable mean NR while reducing computation by approximately 26%. Despite its larger parameter count, WaveNet-VNN achieves lower mean A-weighted NR than the proposed method over 30–350 Hz. Its stronger attenuation at 20–30 Hz (Fig. 3) may reflect benefits of nonlinear modeling, but these gains fall outside the evaluation band.

**Table 1.** The evaluated controllers are compared in terms of mean NR, estimated computational cost, and model parameter count.

| Model | Mean NR (dB(A)) | Cost (kMAC/sample) | Params. |
|---|---|---|---|
| 256-tap FIR | 7.47 | 24.6 | 24.6k |
| 512-tap FIR | 7.49 | 49.2 | 49.2k |
| FxLMS [2] | 0.58 | 64.5 | 4.6k |
| WaveNet-VNN [14] | 6.79 | 493.6 | 493.6k |
| FG-DNN [18] | 6.03 | 900.5 | 900.6k |
| Proposed method | 7.53 | 18.1 | 48.4k |

Using weights from the same jointly trained checkpoint, full causal routing outperformed every fixed expert combination in mean NR (Table 2). The gain over the best fixed combination, $W_0 + \Delta W_a$ , was 0.26 dB. Expert $\Delta W_c$ accounted for 68.15% of routing decisions yet using it exclusively yielded 7.12 dB(A) mean NR, compared with 7.53 dB(A) for full routing. Relative to the baseline FIR alone, expert compensation reduced maximum OOB amplification from 33.47 to 3.09 dB, while mean NR changed from 7.78 to 7.53 dB(A). This reflects a trade-off between in-band attenuation and OOB suppression.

**Table 2.** Mean NR and maximum OOB amplification of the baseline FIR, fixed expert combinations, and full causal routing.

| Configuration | Mean NR (dB(A)) | Max. OOB (dB) |
|---|---|---|
| $W_0$ | 7.78 | 33.47 |
| $W_0 + \Delta W_a$ | 7.27 | 3.06 |
| $W_0 + \Delta W_b$ | 6.76 | 4.26 |
| $W_0 + \Delta W_c$ | 7.12 | 3.16 |
| $W_0 + \Delta W_d$ | 6.85 | 3.09 |
| Full causal routing | 7.53 | 3.09 |

The benefit of routing depends on matching compensation signals to the current input. The proposed method is evaluated here with fixed parameters. Future work will investigate FxLMS-type updates of the baseline FIR or selected expert coefficients under changing acoustic conditions [2].

## 4. CONCLUSIONS

The proposed method combines baseline FIR filtering with MRF-MoE compensation for multichannel ARNC. The optimization framework comprises networked Wiener initialization, graph-regularized expert and router training, and joint refinement. In tests with fixed secondary paths, the proposed method achieved 7.53 dB(A) mean NR while reducing estimated computation by 96.3% relative to WaveNet-VNN [14]. It achieved comparable mean NR to a 256-tap FIR with approximately 26% less computation. Using weights from the same jointly trained checkpoint, full causal routing outperformed every fixed expert combination in mean NR.

## 5. REFERENCES

[1] W. Chen, J. Tao, S. Wang, K. Chen, and H. Zou, "Review on active road noise cancellation technologies for vehicles," (in Chinese), *Acta Acust.*, vol. 50, no. 6, pp. 1380–1394, 2025, doi: 10.12395/0371-0025.2025309.

[2] S. M. Kuo and D. R. Morgan, "Active noise control: a tutorial review," *Proc. IEEE*, vol. 87, no. 6, pp. 943–975, 1999, doi: 10.1109/5.763310.

[3] J. Cheer and S. J. Elliott, "Multichannel control systems for the attenuation of interior road noise in vehicles," *Mech. Syst. Signal Process.*, vol. 60–61, pp. 753–769, 2015, doi: 10.1016/j.ymssp.2015.01.008.

[4] Z. Zhou, S. Chen, H. Li, and Y. Cai, "Delayless partial subband update algorithm for feed-forward active road noise control system in pure electric vehicles," *Mech. Syst. Signal Process.*, vol. 196, 2023, Art. no. 110328, doi: 10.1016/j.ymssp.2023.110328.

[5] L. Zhu, X. Qiu, D. Mao, S. Wu, and X. Zhong, "Efficient segment-update block LMS-Newton algorithm for active control of road noise," *Mech. Syst. Signal Process.*, vol. 198, 2023, Art. no. 110436, doi: 10.1016/j.ymssp.2023.110436.

[6] X. Li *et al.*, "A computationally efficient multichannel feedforward time–frequency-domain adjoint least mean square algorithm for active road noise control," *Appl. Acoust.*, vol. 231, 2025, Art. no. 110441, doi: 10.1016/j.apacoust.2024.110441.

[7] S. Lian *et al.*, "An online decoupling-whitening frequency domain filtered-error least mean square algorithm for active road noise control," *J. Acoust. Soc. Am.*, vol. 156, no. 2, pp. 1413–1424, 2024, doi: 10.1121/10.0028312.

[8] S. Lian *et al.*, "A coherence-based robust frequency-dependent variable step size method for active road noise control," *J. Acoust. Soc. Am.*, vol. 158, no. 2, pp. 1254–1267, 2025, doi: 10.1121/10.0038751.

[9] J. Xue, K. Wen, K. Chen, H. Zou, and J. Lu, "Performance comparison of adaptive active control algorithms for automotive road noise," (in Chinese), *Acta Acust.*, vol. 50, no. 6, pp. 1577–1587, 2025, doi: 10.12395/0371-0025.2025332.

[10] M. de Brett, T. Butlin, L. Andrade, and O. M. Nielsen, "Experimental investigation into the role of nonlinear suspension behaviour in limiting feedforward road noise cancellation," *J. Sound Vib.*, vol. 516, 2022, Art. no. 116532, doi: 10.1016/j.jsv.2021.116532.

[11] J. Y. Oh, H. W. Jung, M. H. Lee, K. H. Lee, and Y. J. Kang, "Enhancing active noise control of road noise using deep neural network to update secondary path estimate in real time," *Mech. Syst. Signal Process.*, vol. 206, 2024, Art. no. 110940, doi: 10.1016/j.ymssp.2023.110940.

[12] C. K. Lai, B. Lam, D. Shi, and W.-S. Gan, "Real-time modelling of observation filter in the remote microphone technique for an active noise control application," in *Proc. IEEE Int. Conf. Acoust., Speech, Signal Process. (ICASSP)*, 2023, pp. 1–5, doi: 10.1109/ICASSP49357.2023.10095752.

[13] H. Zhang and D. Wang, "Deep MCANC: a deep learning approach to multi-channel active noise control," *Neural Netw.*, vol. 158, pp. 318–327, 2023, doi: 10.1016/j.neunet.2022.11.029.

[14] L. Bai *et al.*, "WaveNet-Volterra neural network for active noise control: a fully causal approach," *Mech. Syst. Signal Process.*, vol. 241, 2025, Art. no. 113486, doi: 10.1016/j.ymssp.2025.113486.

[15] Z. Zhang *et al.*, "A causal time-frequency Mamba architecture with Volterra nonlinear modeling for multi-channel automotive road noise control," *Eng. Appl. Artif. Intell.*, vol. 181, 2026, Art. no. 115517, doi: 10.1016/j.engappai.2026.115517.

[16] D. Shi, W.-S. Gan, B. Lam, and S. Wen, "Feedforward selective fixed-filter active noise control: algorithm and implementation," *IEEE/ACM Trans. Audio, Speech, Lang. Process.*, vol. 28, pp. 1479–1492, 2020, doi: 10.1109/taslp.2020.2989582.

[17] Z. Luo, D. Shi, X. Shen, J. Ji, and W.-S. Gan, "Deep generative fixed-filter active noise control," in *Proc. IEEE Int. Conf. Acoust., Speech, Signal Process. (ICASSP)*, 2023, pp. 1–5, doi: 10.1109/ICASSP49357.2023.10095205.

[18] L. Bai, Y. He, X. Nan, K. Chen, and J. Lu, "Feedback-guided DNN-based controller fusion for robust fixed-parameter active noise control," 2026, arXiv:2608.14061, doi: 10.48550/arxiv.2608.14061.

[19] L. Bai *et al.*, "An adaptive deep neural network for active road noise control," *J. Acoust. Soc. Am.*, vol. 159, no. 4, pp. 3674–3685, 2026, doi: 10.1121/10.0043581.

[20] Fareedha, Vasundhara, A. Kar, and M. G. Christensen, "Joint deep secondary path estimation and adaptive control for active noise cancellation," in *Proc. IEEE Int. Conf. Acoust., Speech, Signal Process. (ICASSP)*, 2026, pp. 15177–15181, doi: 10.1109/ICASSP55912.2026.11461474.

[21] Z. Yang, L. Rao, Z. Luo, D. Shi, Q. Huang, and W.-S. Gan, "Co-initialization of control filter and secondary path via meta-learning for active noise control," in *Proc. IEEE Int. Conf. Acoust., Speech, Signal Process. (ICASSP)*, 2026, pp. 15297–15301, doi: 10.1109/ICASSP55912.2026.11463219.

[22] Z. Luo, D. Shi, and W.-S. Gan, "A hybrid SFANC-FxNLMS algorithm for active noise control based on deep learning," *IEEE Signal Process. Lett.*, vol. 29, pp. 1102–1106, 2022, doi: 10.1109/LSP.2022.3169428.

[23] S. Boyd, N. Parikh, E. Chu, B. Peleato, and J. Eckstein, "Distributed optimization and statistical learning via the alternating direction method of multipliers," Found. Trends Mach. Learn., vol. 3, no. 1, pp. 1–122, 2011, doi: 10.1561/2200000016.

[24] N. Parikh and S. Boyd, "Block splitting for distributed optimization," Math. Program. Comput., vol. 6, no. 1, pp. 77–102, 2014.

[25] M. Belkin, P. Niyogi, and V. Sindhwani, "Manifold regularization: A geometric framework for learning from labeled and unlabeled examples," J. Mach. Learn. Res., vol. 7, pp. 2399–2434, 2006.

[26] Acoustical Society of China. "Fourth Shenghua Cup Acoustics Technology Competition" (in Chinese). May 29, 2026. Accessed: Sep. 15, 2026. [Online]. Available: https://www.aschina.org.cn/details.html?contentId=851&id=71